\documentclass[sigconf,screen,nonacm]{acmart}

\AtBeginDocument{%
  }

\usepackage{microtype}
\usepackage{xspace}

\newcommand{\eg}{\emph{e.g.},\xspace}

\begin{document}

\title{STraceBERT: Source Code Retrieval using Semantic Application Traces}

\author{Claudio Spiess}
\orcid{0009-0000-5932-7022}
\email{cvspiess@ucdavis.edu}
\affiliation{
	\institution{University of California at Davis}
	\country{USA}
}

\renewcommand{\shortauthors}{Spiess}

\begin{abstract}
	Software reverse engineering is an essential task in software engineering and security, but it can be a challenging process, especially for adversarial artifacts. To address this challenge, we present STraceBERT, a novel approach that utilizes a Java dynamic analysis tool to record calls to core Java libraries, and pretrain a BERT-style model on the recorded application traces for effective method source code retrieval from a candidate set. Our experiments demonstrate the effectiveness of STraceBERT in retrieving the source code compared to existing approaches. Our proposed approach offers a promising solution to the problem of code retrieval in software reverse engineering and opens up new avenues for further research in this area.
\end{abstract}


\maketitle

\section{Introduction}
Software reverse engineering is a critical task in the field of software engineering and security. It involves the reconstruction of the source code from the compiled binary code to understand the functionality and vulnerabilities of the software. The task becomes more challenging in an adversarial setting, where the binary may be intentionally constructed to hinder analysis. A common example is ransomware, which can be expected to be obfuscated and invoke antianalysis techniques. A reverse engineer wants to understand how the malware works, so they can discover weaknesses, capabilities, and vulnerabilities used. Yet, understanding behavior from millions of assembler instructions is untenable, and practitioners have developed decompiler tools such as Ghidra for native binaries, or Procyon, CFR, and Fernflower for Java bytecode. However, these tools are sensitive to obfuscation techniques, and the usefulness of their output is limited in many such cases.

We hypothesize that while any binary can obfuscate their control flow, variable names, etc, they cannot hide their system calls without a rootkit. Such system calls can be recorded on Unix systems with the \texttt{strace} command. We hypothesize that a complex enough sequence of system calls can be used to retrieve the source code that initiated them, using an embedding model and vector database to retrieve the most similar known sequence of calls with source code. Due to the challenge of aligning method boundaries and source code with system call sequences, we adapt this problem to the Java ecosystem. We utilize a Java dynamic analysis tool \textsc{Jackal}, that instruments bytecode on the fly and records events such as method calls, exits, and field modifications.

We draw an analogy between system calls and calls into Java core libraries i.e. \texttt{java.*}. Using \textsc{Jackal}, we trace the test suites of a set of open source Java projects, record their true source code, train a BERT-style model on the sequences of Java calls, embed all sequences, and then evaluate the similarity of proposed source code candidates for a holdout set. Through our experiments, we demonstrate the effectiveness of our proposed model in retrieving the source code compared to existing baselines.

\section{Background and Related Work}
We searched for similar works applying dynamic analysis for obtaining or augmenting source code, decompilation with ML augmentation, and clone detection. There is a line of work that uses ML methods to improve decompiled code~\cite{summarise_decompiled_binaries, dire}. Another line of work looks at binary clone detection \& binary diffs~\cite{deepbindiff, binclone, binmatch}. Finally, a line of work looks at embedding binary code~\cite{asm2vec}. We found only some work~\cite{trex} which uses dynamic analysis techniques in a similar fashion as this work.

\section{Approach}
\begin{figure}[h]
	\centering
	\includegraphics[width=0.33\textwidth]{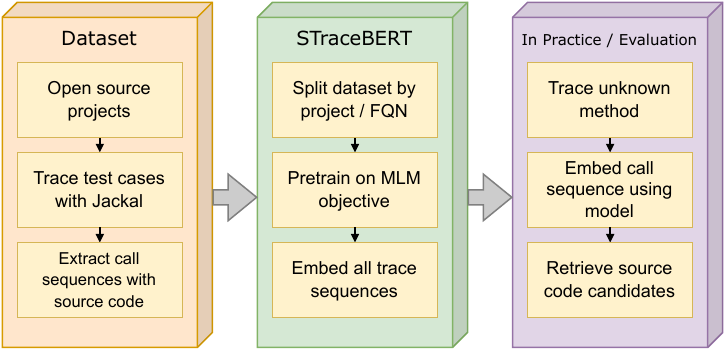}
	\caption{Overview of the STraceBERT approach.}
	\label{fig:overview}
\end{figure}

As our approach is novel, we begin by constructing the Java Trace Dataset (JTD)~\cite{jtd} as outlined in~\autoref{fig:overview}. The JTD consists of Java application traces obtained from a variety of open-source projects. Each trace in the JTD dataset is a sequence of method invocations that represents the execution of a test case. To produce this dataset, we gather a list of top Java projects on GitHub by number of stars, and trace 30 projects' test suites and record call sequences, call sequences \emph{with} call boundaries, source code, and various metadata such as maximum call graph depth at a method level, excluding the test case method itself. A high-quality test suite is optimal for constructing a candidate set for retrieval, but once constructed, binaries being reverse engineered do \emph{not} require a test suite.

We remove all duplicate call sequences to ensure there are no exact duplicate call sequences between traces. We note that for each method, an average of 12 traces were recorded. For such cases, the call sequence and thus call sequence with boundaries are unique, but the source code is the same. We then pick four projects for which the traces act as the candidate set. We split the data into three sets: candidates, for which traces of a particular method by fully-qualified name (FQN) occur \emph{at least once} in candidate project traces, and queries, which consists of traces for which the FQN was encountered \emph{only} in non-candidate projects. As our test sets, we create the \textsc{With Libraries} set by randomly sampling 10k traces from the candidate set without replacement, and \textsc{Without Libraries} by randomly sampling 10k traces from the queries set. This leaves us with a candidate set and two query sets for evaluation: one where we assume the binary under examination uses common libraries such as \textsc{Apache Commons}, and one that doesn't. This provides an upper and lower bound on expected performance.

For pre-training the STraceBERT models, we use all traces in the candidate set, with or without associated source code as the training set. We use a sample of 10k traces from the training set for validation. An example of a call is \texttt{-> java.lang.String.trim(): java.lang.String}. We represent each call signature by one unique token in our vocabulary, to maximize the number of calls we can ingest in our model, and to reduce vocabulary size. For sequences with call boundaries, method invocations and exits outside of the \texttt{java.*} namespace are demarcated by the special tokens \texttt{[CALL]} and \texttt{[EXIT]} respectively. We do not include the name of the called method to avoid data leakage, as the method names would be obfuscated in practice.

Our model is an adapted BERT~\cite{bert} model with a custom tokenizer designed for call sequences. We retain the default hyperparameters, except the attention window size increase to 768, and due to compute limitations, we decrease the following hyperparameters from the default by 33\%: intermediate size of 2048, hidden size of 512, 8 attention heads, and 6 hidden layers. Hyperparameter optimization is left to future work. We first pretrain the model on call sequences from the JTD, using standard Masked Language Modeling (MLM), and default mask probability of 15\%. We train for 400 epochs. Once the model is pretrained, it produces embeddings for each trace in the corpus. These embeddings are used for source code retrieval.

\section{Evaluation and Results}
Our goal is to retrieve candidate snippets of code for a reverse engineer. We presume that presenting five examples to the reverse engineer will allow them to better understand the method they are working with. To evaluate how useful a retrieved snippet is, we measure code similarity between the query ground-truth source code versus each retrieved candidate snippet. For this purpose, all metrics are presented in the form of \textbf{top@\textit{k}}, meaning the average maximum similarity score between the ground-truth source code and the top $k$ retrieved candidates for all queries \eg top@5 would be the maximum CodeBLEU~\cite{codebleu} score of the first five candidates retrieved for each query, averaged across all queries. We utilize the CodeBLEU code similarity measure due to its high correlation with human scores of up to $R^{2} = 0.9548$.~\cite{codebleu} For each trace in our evaluation sets \textsc{With Libraries} \& \textsc{Without Libraries}, we retrieve five nearest-neighbors from our candidate set using FAISS~\cite{faiss} and the trace embeddings from our models. We then calculate and record the similarity metric.

\begin{table}[!ht]
	\caption{Retrieval performance in terms of average maximum CodeBLEU@k, \textsc{With Libraries} and \textsc{Without Libraries}.}
	\label{table:results}
	\begin{tabular}{lcccc}

		               & \multicolumn{2}{c}{\small{\textsc{$+$ Libraries}}} & \multicolumn{2}{c}{\small{\textsc{$-$ Libraries}}}                                   \\
		\cmidrule(lr){2-3}\cmidrule(lr){4-5}
		Top@$k$        & @1                                                 & @5                                                 & @1             & @5             \\

		STraceBert     & 75.09                                              & \textbf{90.88}                                     & 21.72          & 28.99          \\
		$+$ Boundaries & 86.26                                              & \textbf{93.48}                                     & \textbf{22.08} & 29.55          \\
		\hline
		Random         & 18.03                                              & 29.37                                              & 16.95          & 26.50          \\
		Codex          & 29.85                                              & 38.39                                              & \textbf{22.08} & \textbf{31.45} \\
		BM25           & 37.57                                              & 61.11                                              & 17.70          & 23.44          \\
		\hline
		CodeBERT       & 98.71                                              & 98.77                                              & 31.75          & 33.57
	\end{tabular}
\end{table}

We present our results in~\autoref{table:results}. As a reference, we provide a random baseline where the candidates are picked randomly, BM25~\cite{bm25}, and Codex~\cite{chen2021evaluating}. For Codex, we sample 100 queries from each query set, and create an 8-shot prompt following~\cite{nashid2023retrieval} by finding the nearest neighbors of the trace query in the candidate set, filtered to have less than 100 Java calls and less than 300 characters of source code due to context window limitations. We also compare a strawman retrieval using the CodeBERT~\cite{codebert} embedding of the source code. Due to source code overlap but lack of \emph{trace sequence} overlap between \textsc{With Libraries} and candidate set, retrieval via CodeBERT finds exact matches. Using trace embeddings, we find for the average query \emph{with} call boundaries, the first candidate has an average CodeBLEU value of 86.26 \emph{vs.} 75.09 \emph{without} call boundaries. Performance differences decrease @5, with 93.48 \emph{vs.} 90.88 respectively. Further analysis revealed that CodeBLEU values were not a result of exact matches, suggesting similar code produces similar sequences of calls. This presents strong evidence that trace embeddings can be used to find similar code.

\section{Conclusion}
In conclusion, the results of our experiments demonstrate the effectiveness of the STraceBERT approach for the code retrieval task in software reverse engineering. We find that embedding Java application traces using a BERT-style model allows for effective source code retrieval, outperforming traditional information retrieval approach BM25, and modern few-shot prompting of Codex in a known-libraries scenario. Future work includes examining robustness of our approach against obfuscation, a security-related evaluation, and further exploration of the Java Trace Dataset.

\bibliographystyle{ACM-Reference-Format}
\bibliography{bibliography}

\end{document}